Low probability Bid-Bax reaction generates heterogeneity in apoptosis resistance of cancer and cancer stem cells

Subhadip Raychaudhuri

Department of Biomedical Engineering, Biophysics Graduate Group, Graduate Group in Applied Mathematics, University of California, Davis, California 95616

Address Correspondence: raychaudhuri@ucdavis.edu

**Abstract**

Variability in the tumorigenic potential among cancer cells within a tumor population is an unresolved fundamental issue in cancer biology. It is important to know whether cancer cells with higher tumorigenic potential, such as cancer stem cells, are only a small subpopulation. We attempt to address the question of variability in tumorigenic potential based on the heterogeneity in apoptosis resistance of cancer cells. We use stochastic differential equations and kinetic Monte Carlo simulations to explore the mechanisms that generate cell-to-cell variability in apoptosis resistance of cancer cells. In our model, a simplified scheme of apoptosis signaling reactions is developed focusing on the proapoptotic Bid-Bax reaction and its inhibition by Bcl-2 like antiapoptotic proteins. We show how a combination of low probability Bid-Bax reaction along with overexpressed reactant molecules allows specific killing of cancer cells, especially under targeted therapy such as Bcl-2 inhibition. This low probability Bid-Bax reaction protects normal cells from accidental apoptosis but generates cell-to-cell stochastic variability in apoptotic activation of cells equipped with overexpressed Bid and Bax molecules. We further demonstrate that cellular variations in Bcl-2 / Bax ratio, within a cancer cell population, can affect the intrinsic fluctuations arising from the stochastic Bid-Bax reaction and thereby provides a mechanism for origin of cells with higher tumorigenic potential. We discuss the implications of our results for cancer therapy, such as, optimal strategies to minimize stochastic fluctuations in cancer cell death.

**Introduction**

Variability in the tumorigenic\leukaemogenic potential among cancer cells has led to the hypothesis of cancer stem cells (1,2). Existence of such specialized class of cancer cells is an unresolved fundamental issue in cancer biology that has significant implications for cancer treatment (1-4). Previous experimental studies have indicated that only rare population (< 0.1%) of cancer cells possess the ability to form tumors when transplanted into immunodeficient NOD / SCID mice (3,5-7). Recent in vivo studies, using highly immunocompromised NOD / SCID IL-2rg$^{-/-}$ mice, however, shows that a large fraction (~ 25%) of melanoma cancer cells could form tumors (4). Thus one is faced with two related questions in cancer biology: (i) how to characterize the subset of cancer cells that can be labeled as cancer stem cells, and (ii) what fraction of cancer cells has the tumorigenic potential. We attempt to answer these fundamental questions on the basis of heterogeneity in the apoptosis resistance of cancer cells.

Mathematical and computational approaches are well suited to clarify the basis of cell-to-cell stochastic variability in apoptotic activation through the intrinsic (mitochondrial) pathway of cancer cells. In our proposed computational model, the basis of disparity in tumorigenic potential of cancer cells lies in the cell-to-cell variability in apoptotic resistance of cancer cells. Part of this cell-to-cell variability originates from inherent stochastic fluctuations (8-12) in apoptotic activation of cancer cells. Such stochastic activation would support the possibility that all cancer cells have similar but low proliferating potential in *in vitro* colony forming assay or *in vivo* transplantation assay (1). Cell-to-cell variability in apoptosis resistance can also arise from cellular variability in concentrations of key apoptotic regulators such as Bcl-2 (11-13). This type of variability might help isolate a phenotypically distinct population such as CD34+CD38- leukemic stem cells and supports a model where only a small subset of cancer cells posses extensive proliferating potential (5-7). Computational study of the intrinsic pathway of apoptosis allows us to estimate the contributions of both inherent stochastic fluctuations and cellular variations in protein levels in generating cell-to-cell variability in apoptosis resistance of cancer cells.

It has been shown that BH3 mimetic molecules such as HA14-1 or ABT-737 can selectively kill cancer cells leaving normal cells unharmed (11,12,14-20). Such increased potential for apoptotic activation in cancer cells can be explained by high expression levels of Bid-like (BH3 only activators) and Bax-like (multidomain proapoptotic) molecules in cancer cells and assuming that Bid-Bax reaction is a low probability event. In this study, we focus on the pre-mitochondrial signaling module to elucidate how cancer cells, but not normal cells, have a greater chance of getting activated through the Bid-Bax reaction. Increased expression of apoptotic inhibitors, such as Bcl-2 proteins (21-23), would prevent Bid-Bax reaction and thus protect cancer cells from apoptotic death. In addition, a fraction of cancer cells, such as the CD34+CD38- phenotype of leukemic stem cells (5-7), can become particularly resistant to apoptotic stress due to pronounced overexpression of Bcl-2 like antiapoptotic proteins. Once the apoptotic inhibition of Bcl-2 proteins is overwhelmed by BH3 mimetic molecules, cancer cells become vulnerable to apoptotic death and more so than normal cells. Even though a low probability Bid-Bax reaction and its Bcl-2 inhibition can be postulated, lack of measurements of Bid-Bax reaction constants and Bid/Bax/Bcl-2 protein levels makes any quantitative study of apoptotic activation in cancer cells difficult. We first use a computational approach to estimate a possible range in Bid-Bax reaction constant that is further used in subsequent simulations.

We study the intrinsic pathway of apoptosis in which the activation is initiated through the regulatory loop created by Bid-Bax-Bcl-2. A stochastic differential equation based model elucidates the inherently stochastic nature of Bid-Bax reaction. A detailed kinetic Monte Carlo model is then used to simulate both pre- and post-mitochondrial signaling modules in the intrinsic apoptotic pathway. We demonstrate how a low probability of Bid-Bax reaction can protect normal cells from accidental activation and selectively kill cancer cells having overexpressed BH3 only proteins. This low probability Bid-Bax reaction also provides a mechanism for the non-genetic cell-to-cell stochastic variability in apoptotic activation. We demonstrate how increasing levels of Bcl-2 / Bax ratio can amplify the inherent stochastic fluctuations arising from Bid-Bax reaction and thus provides a mechanism for origin of cells with higher tumorigenic potential. We further discuss possible strategies to eliminate large cell-to-cell stochastic fluctuations in the intrinsic apoptotic pathway of cancer cells.

**Method and Results**

We develop and study a simplified network model of apoptotic signaling (24) that is triggered by the direct Bid-Bax reaction (Figure 1). In our previous computational studies we used Bid-tBid conversion reaction to activate the intrinsic pathway of apoptosis (8,10-12). In this study, Bcl-2 inhibition by itself can lead to the activation of the pre-mitochondrial signaling module through direct Bid-Bax reaction. In our model, Bcl-2 represents all the Bcl-2 family proteins with anti-apoptotic properties. Bid is representative of BH3 only activators that can directly react with Bax-like multidomain proapoptotic molecules (captured by Bax in our model). We do not explicitly simulate BH3 only sensitizers (such as Bad or Bik) the effect of which can be simulated by varying Bcl-2 concentrations.

Given the lack of knowledge of Bid-Bax reaction constants, we use our simulations to estimate a low probability of reaction ($K_A \sim 2 \times 10^5$ $M^{-1}$) that can activate the pathway when Bid and Bax are significantly over-expressed, as found in cancer cells, but would prevent accidental activation in normal cells. An order of magnitude variation in the affinity constant ($K_A$) is possible and can still capture many of the qualitative results. Such low probability of reaction generates inherent stochastic fluctuations, and can cause large cell-to-cell variability in apoptotic activation as could be assessed by variability in cytochrome c release. Anti-apoptotic action of Bcl-2 like inhibitors rely on its binding to Bid and Bax, and thus overexpression of Bcl-2 proteins is a mechanism of apoptotic resistance of cancer cells that are equipped with higher concentrations of Bid and Bax molecules. Bid-Bcl-2 reaction constant is also assumed to be low ($K_A \sim 2 \times 10^6$ $M^{-1}$) in proportion to the low activation constant of direct Bid-Bax reaction (compared to tBid-Bax reaction constant whose value is known to be $K_A \sim 2 \times 10^7$ $M^{-1}$). Bid-Bax-Bcl-2 creates a regulatory loop network in the pathway and we study the effect of variation in concentrations of all three molecules that constitutes the loop structure. Variations in Bcl-2 can capture the effect of post-translational cellular variability in protein level as well as the action of targeted inhibitors such as BH3 mimetic agents. Cytochrome C is released into the cytosol in an all-or-none manner when Bax activation reaches a pre-assigned threshold value (25). We use a stochastic differential equation based model of the pre-mitochondrial signaling module to elucidate the mechanisms of stochastic fluctuations arising from low probability Bid-Bax reactions. A detailed computational study is carried out using kinetic Monte Carlo (MC) simulations of pre- and post-mitochondrial signaling events (8). In the MC model, cytochrome c release leads to cytochorme c – apaf binding and the subsequent formation of multi-molecular complex apoptosome. Low probability of apoptosome formation generates additional stochastic variability in apoptotic activation, but once formed, induces rapid activation of downstream effecter caspases -9 and -3 (8, 26). Activation of caspase-3 is taken to be a final readout of apoptotic death signaling in a cell and we use Monte Carlo simulations to measure the time-course of caspase-3 activation at a single cell level.

Signaling reactions that describe binding-unbinding between the *i*-th pair of molecular reactants ($A^i + B^i \rightarrow C^i$) are modeled by the following set of stochastic differential equations

$$dA^i = dB^i = -dC^i = -k_{on}^i A^i B^i dt - \sqrt{k_{on}^i A^i B^i} dW + k_{off}^i C^i dt + \sqrt{k_{off}^i C^i} dW$$

W is a Weiner process (27, 28). Stochastic equations were derived from the corresponding master equations that describe the signaling reaction of a given type (27). All three molecules in the Bid-Bax-Bcl-2 regulatory loop undergo binding-unbinding reaction with each other that can be described by stochastic equations as shown above. In addition, Bax can get converted to an activated form when dissociating from the Bid-Bax complex. Such catalytic activation reaction is included in the model using a term similar to the dissociation reaction and a similar rate constant. The stochastic differential equation (SDE) that captures low probability Bid-Bax association was solved using a Poisson Runga-Kutta scheme (29). In this method, the number of times a specific reaction channel fires in a given time t is a Poisson random variable with mean $\lambda t$ and variance $\lambda t$, where $\lambda$ is the propensity function. For the Bid-Bax association reaction $\lambda = k_{on}$[Bid][Bax] where [] denotes the concentration of a molecular species. All other reaction SDEs was solved using the standard Euler-Maruyama numerical scheme of solving stochastic differential equations (29, 30). The system is partitioned into subsets of slow and fast reactions in a manner such that the slow Bid-Bax binding reaction is considered with a stepsize of $\tau$ within which all other (fast) reactions are repeatedly solved (29).

Bax was not activated when we used regular concentrations (as found in normal cells): Bid = 20, Bax = 50, Bcl-2 = 45 (molecules). Molecular numbers are estimated for a 1 µm$^3$ volume from known molar concentrations (31) and discussed in detail in our previous work (8,10-12). A two-fold overexpression in Bid and Bax, however, activated Bax, and hence the pre-mitochondrial signaling module, in a stochastic manner. Thus, even when all the cellular parameters remain identical in a cell population, inherent stochastic fluctuations in the Bid-Bax reaction generates large cell-to-cell variability in cytochrome c release. Kinetic reaction rate constants for the low probability Bid-Bax reaction determine the overexpression levels in Bid (and Bax) molecules needed for activation as well as the extent of cell-to-cell stochastic variability in Bax activation. Cytochrome C release, as a result of threshold Bax activation, took longer as increasing number of Bcl-2 is used: 45, 450, and 675 (Bid = 200, Bax = 500, Figure 2). Cell-to-cell stochastic variability in cytochrome c activation, which arises due to low probability of Bid-Bax association reaction, also increases with increasing Bcl-2 levels and remains sensitive to the Bcl-2 / Bax ratio. As the Bcl-2 / Bax ratio exceeds 1 in our model, a significant increase in stochastic variability in cytochrome c release is resulted as can be seen from Figure 2. Very high Bcl-2 levels (> 5 times Bax) leads to strong inhibition of low probability Bid-Bax reaction and thus provides a mechanism for apoptosis resistance of cancer cells. The threshold Bcl-2 level (to inhibit Bax activation) can be estimated by considering the rate equations of Bid-Bax and Bcl-2-Bax. Assuming Bcl-2-Bax reaction overwhelms available Bax molecules when Bcl-2 / Bax ratio exceeds α (Bcl-2 ≥ α Bax, α is a constant)

$$k_{on}^{Bcl2-Bax}[Bcl2][Bax] \approx k_{off}^{Bcl2-Bax}[Bcl2\text{-}Bax]$$

$$\Rightarrow k_{on}^{Bcl2-Bax}\left(Bcl2^0 - [Bcl2\text{-}Bax]\right)\left(Bax^0 - [Bcl2\text{-}Bax]\right) = k_{off}^{Bcl2-Bax}[Bcl2\text{-}Bax]$$

$$\Rightarrow k_{on}^{Bcl2-Bax}\left(\alpha\left(Bax^0\right) - [Bcl2\text{-}Bax]\right)\left(Bax^0 - [Bcl2\text{-}Bax]\right) = k_{off}^{Bcl2-Bax}[Bcl2\text{-}Bax]$$

assume $Bax^0 - [Bcl2\text{-}Bax] \sim 1$ (small)

$$k_{on}^{Bcl2-Bax} Bax^0 (\alpha - 1) \approx k_{off}^{Bcl2-Bax}[Bcl2\text{-}Bax]$$

$$\Rightarrow k_{on}^{Bcl2-Bax}[Bcl2\text{-}Bax](\alpha - 1) \approx k_{off}^{Bcl2-Bax}[Bcl2\text{-}Bax]$$

$$\Rightarrow (\alpha - 1) \sim \frac{k_{off}^{Bcl2-Bax}}{k_{on}^{Bcl2-Bax}} = \frac{2 \times 10^{-2}}{0.33 \times 10^{-2}}$$

$$\Rightarrow \alpha \sim 7$$

Given the time-scale over which we solve the signaling reactions, Bcl-2 / Bax ~ 5 completely blocked Bax activation and cytochrome c release (over a wide range of Bid and Bax overexpressions). Thus variation in Bcl-2 levels can be a mechanism used by cancer cells to create heterogeneity, such as cancer stem cells, in apoptotic resistance within a tumor. A ten-fold overexpression of Bid and Bax, but keeping the Bcl-2 level at its normal, diminished cell-to-cell stochastic variability in a significant manner. Such a result can be seen from comparing the deterministic and stochastic terms in the SDE for Bid-Bax reaction.

$$k_{on}^{Bid-Bax}[Bid][Bax] \sim \sqrt{k_{on}^{Bid-Bax}[Bid][Bax]}$$

Assuming $[Bid] \approx [Bax] \sim x$

$$x \sim 1/\sqrt{k_{on}^{Bid-Bax}} \sim 500$$

Thus one could observe a stochastic to deterministic transition within the intrinsic pathway of cancer cells when Bcl-2 levels are low or inhibited in a significant manner. One can compare the Bcl-2–Bax binding / unbinding reaction terms and estimate that the ratio Bax / Bcl-2 ~ 6 (as done above) is sufficient to overcome the Bcl-2 inhibition of Bid-Bax reaction. Thus, when Bax / Bcl-2 ratio exceeds a threshold value, as well as the product of Bid and Bax concentrations is higher than $1/\sqrt{k_{on}}$ for the Bid-Bax association reaction, one expects a stochastic to deterministic transition in the release of mitochondrial cytochrome c, induced by the pre-mitochondrial signaling activation. Such a stochastic to deterministic transition will be unique to apoptotic activation in cancer cells and thus it can play a crucial role in developing a successful strategy for killing cancer cells by targeting the intrinsic pathway. Otherwise, inherent stochastic variability in the activation of pre-mitochondrial signaling module is a characteristic feature of cancer cell death, and, partly explains the variability in apoptotic vulnerability among cancer cells as proposed by cancer stem cells (1, 2). Variability in the amounts of Bid/Bax/Bcl-2, as may exist among cancer cells of a clonal population, further modulates the susceptibility to apoptotic activation and its stochastic variability, and thus provides additional mechanism for the generation of cancer stem cells.

In our kinetic Monte Carlo simulation scheme, a set of signaling molecules that belong to the pre- and post-mitochondrial signaling modules are initially distributed randomly and uniformly (8). Signaling molecules are then randomly sampled for diffusion and reaction moves with predefined probability constants. Probabilistic parameters used in our simulations are estimated from known values of kinetic and diffusion rate constants (8, 31). Estimation of molecular numbers (~ 0.6 x nano-molar concentration x V $\mu m^3$) for the simulation volume has been discussed in our previous work (8,10-12). For Bid (33.3 nm), Bax (83.3 nm), Bcl-2 (75 nm) levels found in normal cells, a Bid-Bax association constant of 2 x $10^3$ $M^{-1}$ $s^{-1}$ (compare the same for tBid-Bax: 2 x $10^5$ $M^{-1}$ $s^{-1}$), and a dissociation constant same as that of tBid-Bax reaction: $K_{off}$ = 0.02 $s^{-1}$ ($K_A$ = 2 x $10^7$ $M^{-1}$), protect cells from apoptosis. A two-fold increase in Bid and Bax level still did not activate Bax but that would have happened with a ten-fold higher Bid-Bax association constant. The effect of low rate constant Bid-Bax reaction on Bax activation is stronger in the Monte Carlo model than in the SDE model which can be attributed to the additional requirement of spatial encounter of reactant molecules (by diffusive transport) in the Monte Carlo model. We use the value of $K_{on}$ = 2 x $10^3$ $M^{-1}$ $s^{-1}$ for all further simulations. Slight variations in the value of this association constant would not change the qualitative results presented here.

Increasing the level of Bid and Bax by ten-fold but keeping the Bcl-2 concentration at a normal level led to significant Bax activation and cytochorme c release (Figure 3). Bcl-2 over-expression has been observed in a variety of cancer cells and is a marker for poor prognosis (21-23). In CD34+ leukemia cells, possibly a stem cell phenotype, Bcl-2 has been shown to be highly over-expressed (32,33). Keeping the concentrations of Bid and Bax fixed at a ten-fold overexpression, but gradually increasing the levels of Bcl-2, diminished Bax activation. The fraction of cells in which cytochrome c was released, at a given time point, also decreased with increasing concentrations of Bcl-2. Cell-to-cell stochastic variability in the time-scale of cytochrome c release increased with increasing Bcl-2 levels (Figure 3). For very high Bcl-2 concentrations (Bcl-2 / Bax ~ 5), even when Bid and Bax are ten-fold overexpressed, cytochrome c release was significantly inhibited. Formation of post-mitochondrial apoptosomes, modeled as a multimolecular cytochrome c- apaf complex, led to additional cell-to-cell variability in apoptotic activation as can be seen in Figure 4. The percentage of apoptotic cells, as measured by full caspase-3 activation, is shown in Figure 5 for various values of Bcl-2 inhibition and also at different time instants. The effect of Bcl-2 inhibition on caspase-3 activation is more pronounced at early times (~ 3 hours) of signaling. Increasing the overexpression level of Bid and Bax, but keeping the Bcl-2 / Bax ratio fixed, enhanced Bax activation. Thus in cancer stem cells, if high antiapoptotic Bcl-2 levels are accompanied with similar high expressions of Bid and Bax (keeping the Bcl-2 / Bax ratio fixed), inhibiting Bcl-2 levels would actually lead to more pronounced apoptotic death of cancer stem cells than that of regular cancer cells. Bcl-2 inhibition in the presence of high levels of Bid and Bax, as also found in the theoretical analysis of stochastic equations, would significantly diminish cell-to-cell stochastic variability in apoptotic activation of cancer cells. However, if high Bcl-2 expression is also accompanied with a high Bcl-2 / Bax ratio, then alternative strategies, such as removing the BH4 domain or opening up the BH3 domain of Bcl-2 (34) to make it proapoptotic, would selectively kill cancer stem cells. Bid truncation to tBid, could be another potential option in such cases of high Bcl-2/Bax ratio. Thus two important cellular parameters modulate cytochrome c release and as well as cell-to-cell variability in apoptosis resistance: (i) over-expression level of Bax and Bid like BH3 only proteins, (ii) Bcl-2 / Bax ratio. Over-expression of Bax and Bid underlies the specific killing of cancer cells, especially when Bcl-2 inhibition can be removed. Bcl-2 / Bax ratio determines the degree of apoptosis resistance and leads to heterogeneity among cancer cells.

**Discussion**

Recent experiments have conclusively shown the existence of large cell-to-cell variability in apoptotic activation through the intrinsic mitochondrial (type 2) pathway (11-13, 26). All-or-none type behavior in cytochrome c release and capase-9 /3 activation, along with large cell-to-cell variations in the time-scale of such activation events, characterize cell-to-cell fluctuations in apoptotic activation. Molecular mechanism of cell-to-cell variability remains a fundamental question in the biology of apoptosis. Cellular variability in pro- and anti- apoptotic protein levels that act upstream of mitochondrial outer membrane permeabilization (MOMP), coupled with a rapid release of cytochrome c once a threshold apoptotic stress has gathered on mitochondria, can cause such cell-to-cell variability in cytochrome c release and thus downstream activation of caspases (13). However, even when all the cellular parameters such as relevant protein concentrations remain identical in a cell population, inherent stochastic fluctuations in signaling reactions can cause large cell-to-cell stochastic variability (8-10). Our earlier computational studies indicated that inhibition of tBid-Bax reaction by large amount of Bcl-2 (11,12) and low probability of apoptosome formation (8, 26) are two possible mechanisms that can generate such inherent stochastic fluctuations in apoptotic activation of normal and cancer cells (35, 36). In this study, we show that the low probability Bid-Bax reaction is another dominant mechanism for generating inherent stochastic fluctuations in cell death and such a mechanism is highly relevant for cancer cells.

Using a set of stochastic differential equations and kinetic Monte Carlo simulations we elucidated the stochastic nature of Bax activation that occurs through a low probability Bid-Bax reaction. We carried out apoptosis simulations for various values of Bid, Bax, and Bcl-2 proteins and gathered quantitative information about apoptotic activation. Variations in Bid and Bax levels as well as the Bcl-2 / Bax ratio modulates cell-to-cell variability in apoptotic activation arising from stochastic Bid-Bax reaction. Overexpression of Bid like (BH3 only) proteins and Bax proteins, as possibly found in cancer cells, reduces the apoptotic threshold for cancer cells setting them on the verge of apoptotic collapse, but those cells remain protected by highly overexpressed antiapoptotic Bcl-2 like proteins. Our results indicate how a stochastic to deterministic transition in the pre-mitochondrial signaling module can provide unique opportunity for cancer therapy that relies on the apoptotic activation of the intrinsic pathway.

Stochastic activation of single cells due to low probability of Bid-Bax reaction and cellular variation in protein levels would both regulate apoptosis resistance in single cells. Such heterogeneity in apoptosis resistance can be a primary mechanism for generating heterogeneity in a tumor population. Thus single cancer cells will get killed in a stochastic manner in a xeno-transplantation assay and only a fraction of cells can survive the assault of activated immune cells. In a modified xeno-transplantation assay, such as in immunocompromised NOD / SCID IL-2rg$^{-/-}$ mice, less resistant cancer cells having lower Bcl-2 / Bax ratio would survive and produce full-blown tumors. However, only cancer stem cells that are highly resistant to apoptosis, at least partly due to over-expression of Bcl-2 like proteins (and high Bcl-2 / Bax ratio), would survive in NOD / SCID mice. Experimental studies that would undertake quantitative measurement of signaling proteins in the apoptotic pathway of cancer cells, including in cancer stem cells, will further strengthen the results obtained from our computational studies. Previous studies indicated presence of high Bcl-2 levels in cancer stem cell phenotype such as CD34+CD38- cells in leukemia. CD34+CD38- leukemic stem cells were found to be only a small fraction of cancer cells in a variety of leukemia (5-7). In contrast, in KG1a, a very aggressive form of acute myeloid leukemia (AML), over 90% of the cells has the CD34+CD38- phenotype (7). Thus one could expect a high frequency of cells having tumorigenic potential when single KG1a cells are transplanted into NOD / SCID mice and provides an opportunity to test whether any variations in tumorigenic potential would arise from inherent stochastic fluctuations in apoptosis resistance. In general, a combination of the stochastic factor and variations in cellular protein levels would generate significant heterogeneity within a tumor cell population and such heterogeneity would also heavily depend on the cell type under consideration.

**Acknowledgement**
We thank Arjun Nair for carrying out data analysis and plotting. We also thank Somkanya C. Das for help with preparing this article.

**Figures**

Figure 1. Schematic of a simplified network for the intrinsic (mitochondrial) pathway of apoptosis. Bid-Bax reaction activates the pre-mitochondrial signaling module resulting in release of mitochondrial cytochrome c and activation of effector caspases in the post-mitochondrial signaling module.

Figure 2. Cytochrome C release, as a result of threshold Bax activation in the SDE model, is plotted for single cells. The number of Bcl-2 molecules is varied in our simulations; Bcl-2 = 45, 450, and 675 are used (Bid = 200, Bax = 500). Rapid all-or-none type release of cytochrome c is approximated by a sharp spike (step function) and normalized data is shown. Note, the time-scale of cytochrome c release significantly changes as the Bcl-2 / Bax ratio exceeds one.

Figure 3. Single cell activations obtained from Monte Carlo (MC) simulations of the intrinsic pathway of apoptosis. We show the release of cytochrome c (normalized data in sharp-rising approximation) for various values of Bcl-2 = 45, 450, and 900 (Bid = 200, Bax = 500). Data is shown up to $10^8$ MC steps (1 MC step = $10^{-4}$ sec.).

Figure 4. Time course of caspase-3 activation, as readout for downstream apoptotic activation, is shown for single cells. Simulations are run for 5 x $10^8$ MC steps (1 MC step = $10^{-4}$ sec.). Normalized data is shown for various values of Bcl-2 = 45, 450, and 900 (Bid = 200, Bax = 500).

Figure 5. Percentage of apoptotic cells, as measured by full caspase-3 activation, is shown for various values of Bcl-2 inhibition: Bcl-2 = 45, 450, 900, 1350, 1800, and 2250 (left to right). Data is shown for three different time instants: t = $10^8$, 2 x $10^8$, and 5 x $10^8$ Monte Carlo steps. Cells used to estimate apoptotic activation include cells for which single cell data is shown in Figures 3 and 4. Data is not shown if there is no activation (for specific parameter values).

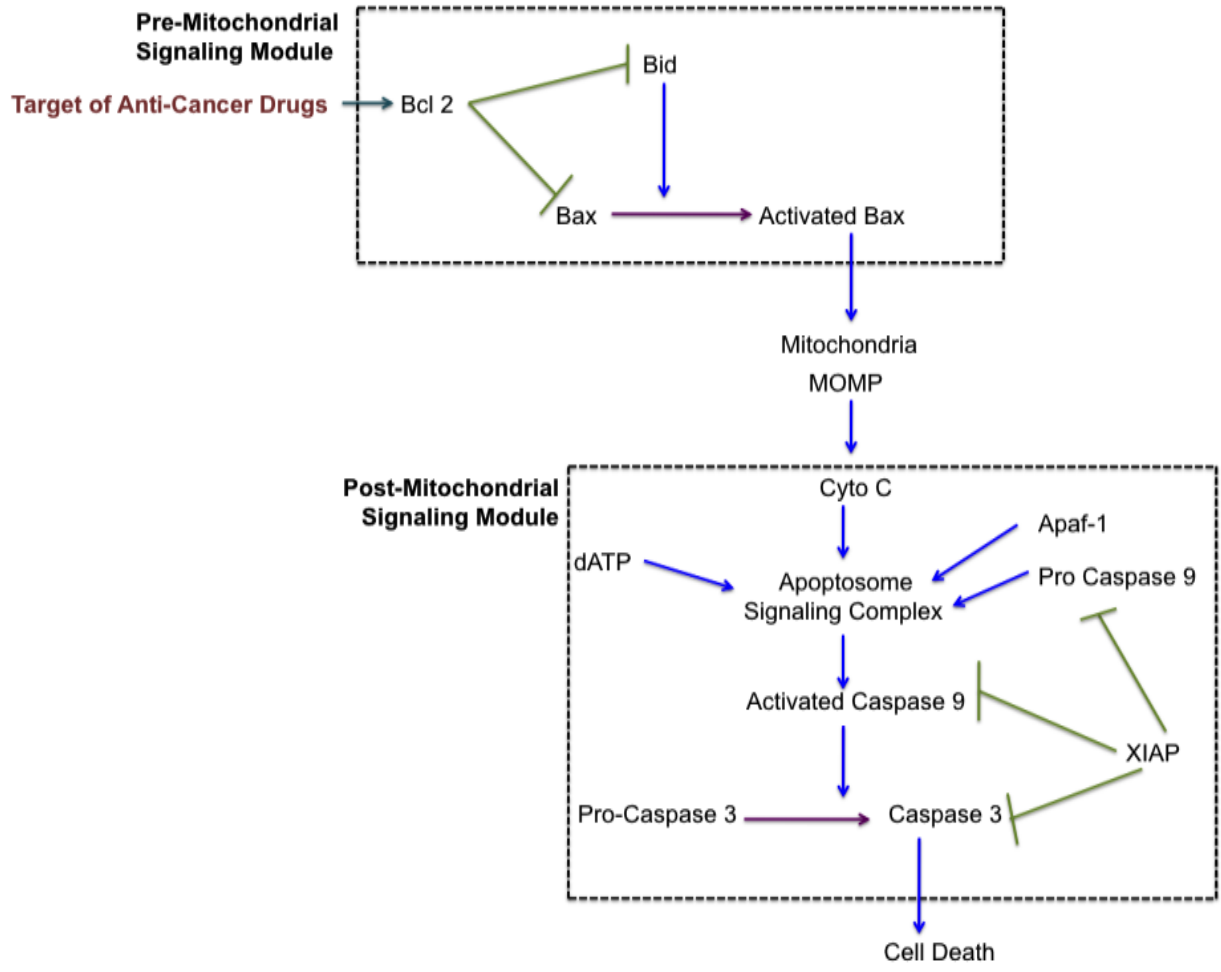

**Figure 1**

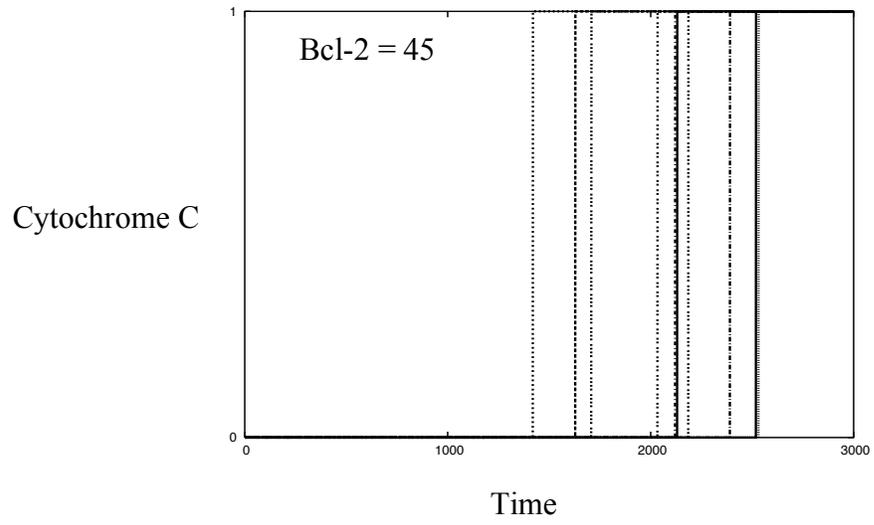
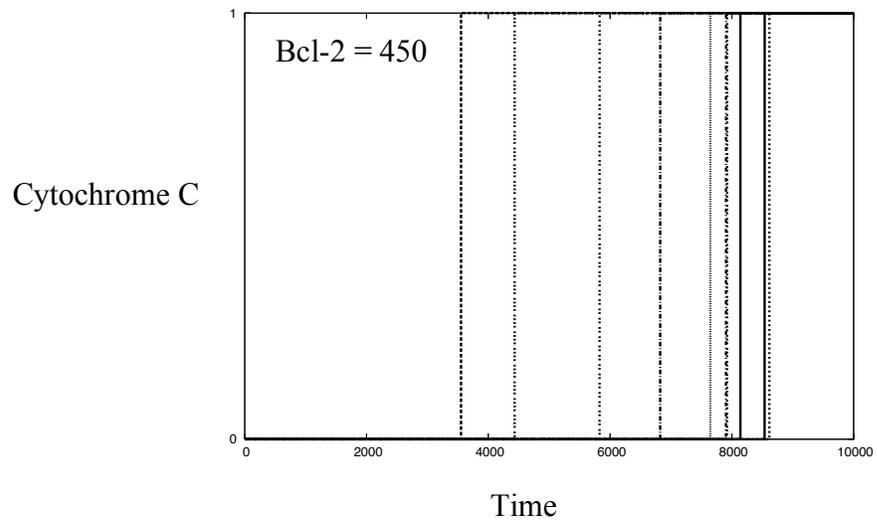
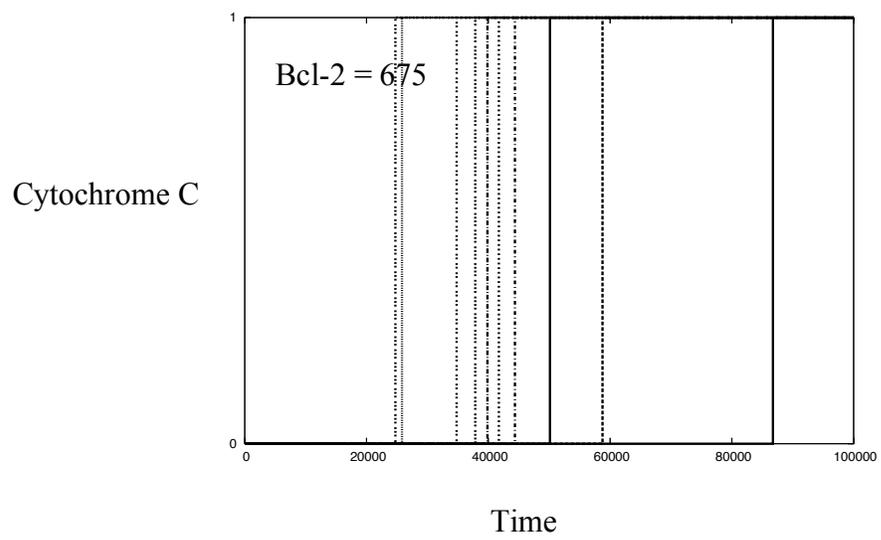

**Figure 2**

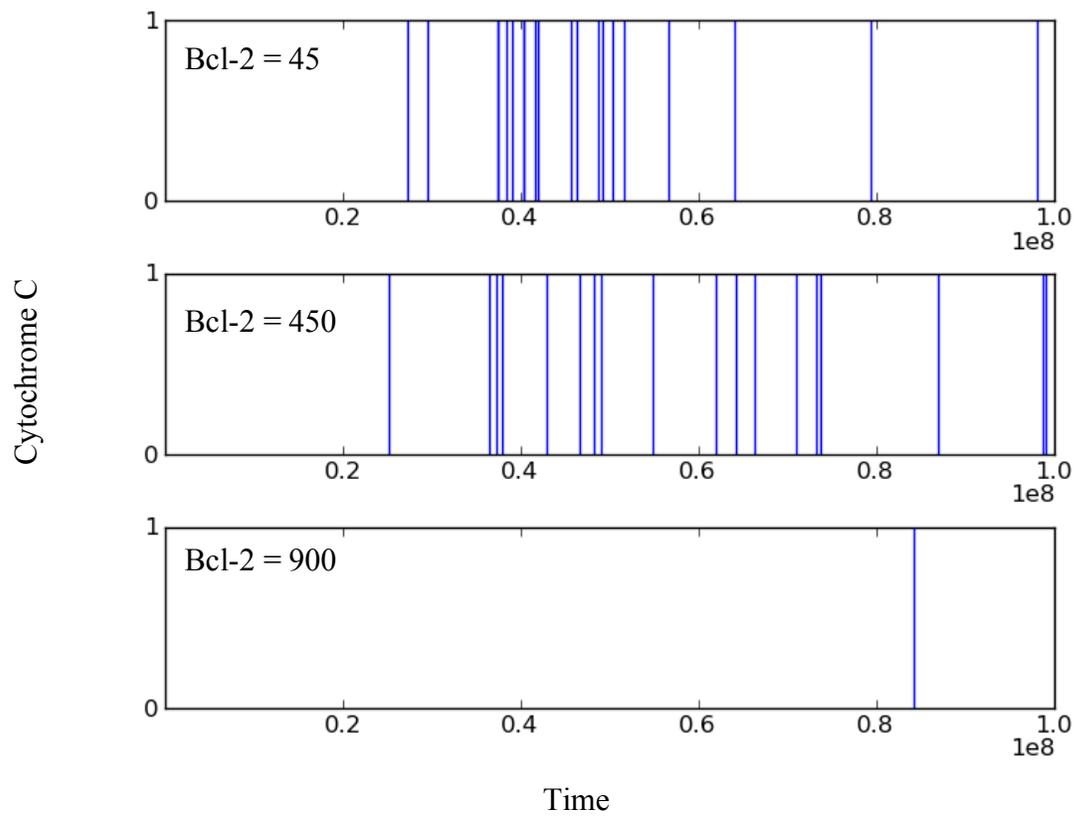

**Figure 3**

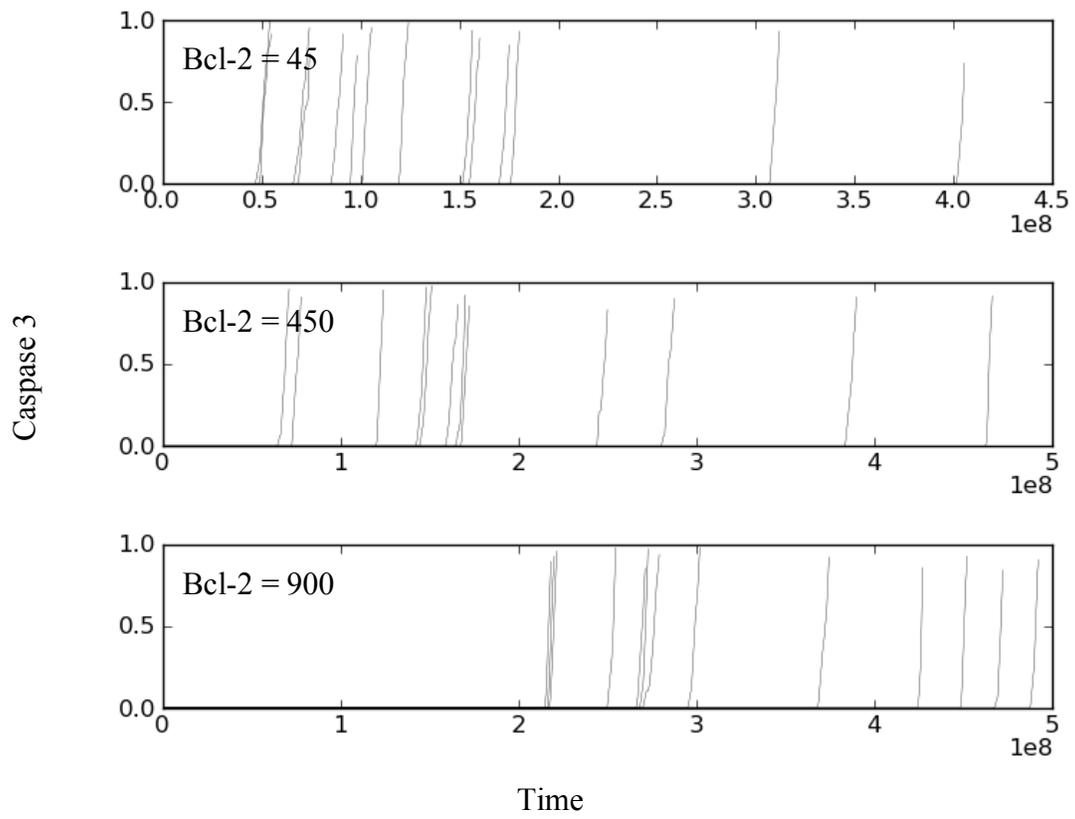

**Figure 4**

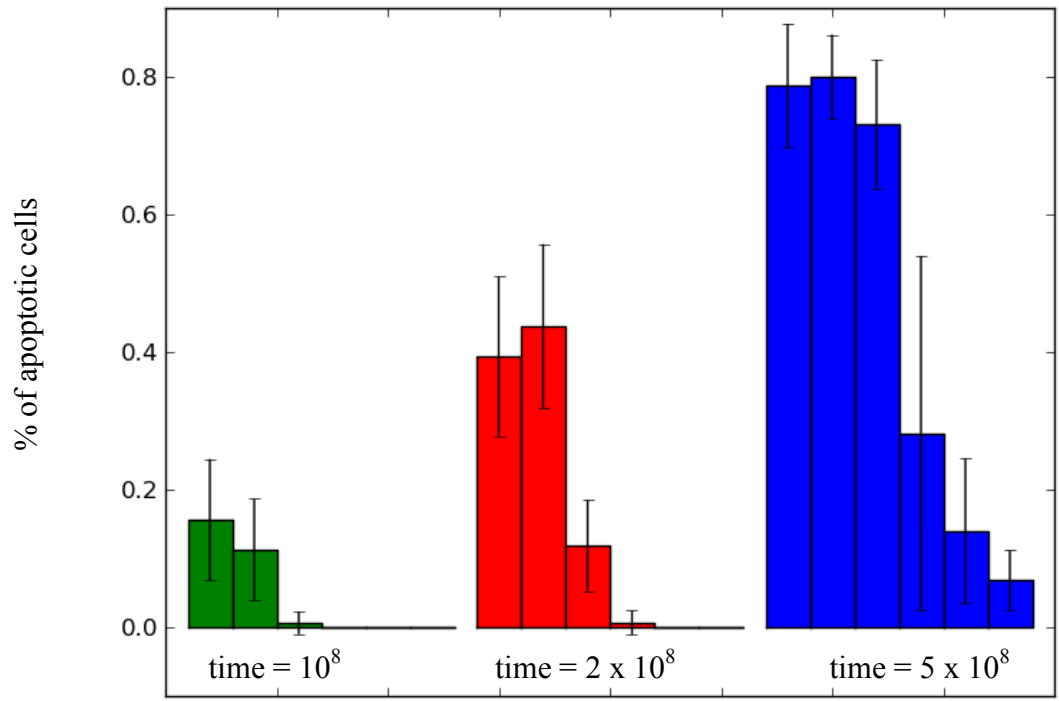

**Figure 5**